\documentclass{aa}  

\usepackage{soul}
\usepackage{threeparttable}
\usepackage{natbib}
\usepackage{diagbox}
\usepackage{slashbox}
\usepackage{multirow}
\usepackage{float}
\usepackage{booktabs}
\usepackage{graphicx,subfigure}
\usepackage{txfonts}
\usepackage[]{hyperref}
\usepackage{appendix}
\newcommand{\urlNewWindow}[1]{\href[pdfnewwindow=true]{#1}{\nolinkurl{#1}}}
\usepackage{color}

\usepackage{epstopdf}
\usepackage[]{sidecap}
\usepackage{multirow}

\begin{document}

%

   \title{Rotational evolution of solar-type protostars during the star-disk interaction phase}

   \author{F. Gallet\inst{1}         
          \and C. Zanni\inst{2}
          \and L. Amard\inst{3}
          }
        \institute{Univ. Grenoble Alpes, CNRS, IPAG, 38000 Grenoble, France 
    \and
         INAF - Osservatorio Astrofisico di Torino, Strada Osservatorio 20, 10025, Pino Torinese, Italy 
     \and 
     University of Exeter, Department of Physics \& Astronomy, Stoker Road, Devon, Exeter, EX4 4QL, UK
        }

\offprints{F. Gallet,\\ email: florian.gallet1@univ-grenoble-alpes.fr}

\authorrunning{Gallet et al.}

   \date{Received -; accepted -}

  \date{Received -; accepted -}
 \abstract
 {The early pre-main sequence phase during which solar-mass stars are still likely surrounded by an accretion disk represents a puzzling stage of their rotational evolution. While solar-mass stars are accreting and contracting, they do not seem to spin up substantially.} 
 {It is usually assumed that the magnetospheric star-disk interaction tends to maintain the stellar rotation period constant (`disk-locking'), but this hypothesis has never been thoroughly verified. Our aim is to investigate the impact of the star-disk interaction mechanism on the stellar spin evolution during the accreting pre-main sequence phases.  }  
 {We devised a model for the torques acting on the stellar envelope based on studies of stellar winds, and we developed a new prescription for the star-disk coupling founded on numerical simulations of star-disk interaction and magnetospheric ejections. We then used this torque model to follow the long-term evolution of the stellar rotation. }
 {Strong dipolar magnetic field components up to a few kG are required to extract enough angular momentum so as to keep the surface rotation rate of solar-type stars approximately constant for a few Myr. Furthermore an efficient enough spin-down torque can be provided by either one of the following: a stellar wind with a mass outflow rate corresponding to $\approx 10\%$ of the accretion rate, or a lighter stellar wind combined with a disk that is truncated around the corotation radius entering a propeller regime.  }
 {Magnetospheric ejections and accretion powered stellar winds play an important role in the spin evolution of solar-type stars. However, kG dipolar magnetic fields are neither uncommon or ubiquitous. Besides, it is unclear how massive stellar winds can be powered while numerical models of the propeller regime display a strong variability that has no observational confirmation. Better observational statistics and more realistic models could contribute to help lessen our calculations' requirements. }
   \keywords{Stars: solar-type -- Stars: magnetic field -- Stars: evolution --  Stars: rotation -- Stars: winds, outflows}
   
   \titlerunning{On the rotational evolution of solar-type protostars}
   \maketitle
   
   \sidecaptionvpos{figure}{c}
   
\section{Introduction}
\label{intro}

Classical T Tauri stars (CTTS) are magnetically active pre-main sequence stars surrounded by an accretion disk \citep{Cameron93,Edwards93,Cameron95,Edwards94,Hartmann98}. Their magnetic field \citep[up to a few kG, see][]{Krull09,Gregory12} has a strong impact on the dynamics and the structure of the stellar environment and can lead to the truncation of the disk and accretion of material along funnel flows down to the stellar surface, the launch of stellar winds along the opened magnetic field lines \citep{MP05b,MP08a,Matt12,Cranmer11,Matt15,Reville15,Johnstone15a,See18,Finley18}, and the ejection of material due to the magnetic star-disk interaction \citep{Shu94,FPA00,Romanova09,Zanni2013}.

Additionally, observations suggest \citep[see][]{Edwards93,Bouvier93,Rebull04,Irwin09a} that while stars contract during the early pre-main sequence (PMS) phase (1-10 Myr) and accrete angular momentum from the disk, they do not seem to noticeably spin up for several Myrs.  
{Indeed, \citet[][and references therein]{Gallet13,GB15} highlight the apparent steady evolution of the three percentiles (90$\rm th$, median, and 25$\rm th$) of the rotational period distributions of stars from 1 Myr to 10 Myr. The fact that the two extreme percentiles remain almost constant in time confirms that the evolution of the rotation is not random but strongly depends on the initial conditions, that is, whether the star is initially in a fast or slow rotating regime.}
{Since this constant rotational phase is comparable to the disk lifetime, it suggests a link between the magnetic star-disk interaction and this observed behaviour.}
{Finally, regardless of the physical origin of this rotational evolution}, these observational constraints suggest that during this early-PMS phase, a large fraction of the angular momentum of the star {needs to be} removed.

{Moreover,} the magnetospheric star-disk interaction scenario is still the main paradigm to interpret the angular momentum evolution of CTTS \citep[see][for a review]{Bouvier14}. The first attempts to model the star-disk angular momentum exchange in CTTS \citep[see e.g.][]{Konigl1991} were based on the \cite{GhoshLamb1979} model, which was originally developed for X-ray pulsars. In this framework, the star can transfer its angular momentum to the disk along magnetic field lines that connect the star with the disk region beyond the corotation radius. This scenario popularized the idea that the disk itself could adjust the stellar rotation at the observed values (disk-locking mechanism). 
In more recent times, the \cite{GhoshLamb1979} model has been revised \citep{MP05a}, thus showing that this mechanism is actually very inefficient since the size of the disk region that is magnetically connected to the star is likely very small and often confined inside the corotation radius \citep{Zanni2013}. As of today, outflows emerging from the star-disk interaction region seem to be the best viable mechanism to extract the excess of stellar angular momentum. \cite{Shu94} proposed that a wide-angle outflow that is launched from a small region located around the corotation radius, the `X-Wind', could be able to extract a sizable fraction of the disk angular momentum at corotation before it falls onto the star, thus at least eliminating the spin-up torque due to accretion. 
Stellar winds, which are possibly powered by accretion \citep[{accretion powered stellar wind, hereafter APSW, see}][]{MP05b}, can directly extract angular momentum from the star.\ Additionally, accurate estimates of the spin-down torque have been computed via direct numerical simulations \citep{MP08a,Matt12,Reville15,Pantolmos2017,Finley18}. Another class of outflows can exploit the stellar magnetosphere that still connects the star to the disk, causing ejections that are intrinsically unsteady and possibly quasi-periodic. If the magnetic moment of the field that threads the disk is aligned with the stellar one, a reconnection X-point forms in the disk midplane, where matter can be uplifted from the disk and accelerated by the stellar rotation along newly opened field lines, thus removing stellar angular momentum \citep[`ReX-wind', ][]{FPA00}. If the stellar and disk magnetic moments are anti-parallel, as in the case where the open disk magnetic field is a leftover of the star-disk interaction, unsteady ejections can arise due to the quadi-periodic process of inflation, opening and reconnection of the closed magnetospheric field lines \citep[{magnetospheric ejections, hereafter MEs,}][]{Romanova09,Zanni2013}. Since these two last outflow types take advantage of magnetic field lines that still connect the star with the disk, they can exchange mass, energy, and angular momentum with both the star and the disk. 

The works by \cite{Gallet13, GB15} and \cite{Amard16}, aimed at comparing theoretical models and observations of the rotational evolution of solar-mass stars, make the hypothesis that during the accreting TTauri phase the stellar rotation is kept constant thanks to a `disk-locking' mechanism, but they did not take into account the actual details of any magnetopsheric star-disk interaction model. The aim of this work is to add self-consistent elements to a physical model of the star-disk interaction and to the evolutionary tracks presented in \citet{Gallet13,GB15} and \citet[][]{Amard16} in order to determine under which physical conditions the stellar rotation period can be kept approximately constant while the protostars are still accreting and contracting. To model the stellar spin evolution we include the spin-down torque exerted by a stellar wind, employing the parametrization proposed by \cite{Matt12}.\ Nevertheless, we present a new prescription for the angular momentum exchange between a star and a surrounding accretion disk based on the numerical magnetohydrodynamic (MHD) simulations of \cite{Zanni2013} that show the impact of {MEs} on the stellar angular momentum evolution. We did not consider a \citeauthor{GhoshLamb1979}-like star-disk magnetic coupling since it has been tested to provide a rather inefficient spin-down torque \cite[see e.g.][]{Matt10}, and we preferred taking into account scenarios that rely on outflows. We did not consider either X-wind \citep{Shu94} or ReX-wind \citep{FPA00} scenarios that are still based on phenomenological models and we preferred to include mechanisms for which self-consistent MHD numerical solutions are currently available.  

The structure of this article is as follows: in Section \ref{model}, we briefly present the angular momentum evolution model together with the different star-disk interaction processes used in this study. The outcome of our rotational evolution models is presented in Section \ref{results}. In Section \ref{disc}, we discuss the implications of these models, in particular on the magnetic field intensity and the kind of star-disk interaction regimes that could provide a torque that is efficient enough to slow down the stellar surface. We finally draw conclusions  about the validity and limitations of these models in Section \ref{conc}. 

\section{The model}
\label{model}

We build upon the model described in \citet{Gallet13} that is dedicated to the study of the angular momentum evolution of solar-type stars {(i.e. stars with solar mass and metallicity)}. In this two-zone model, the stellar convective envelope and the radiative interior are separated entities that exchange angular momentum over a given time-scale, and the stellar internal structure evolves with time. Additionally, during the stellar evolution after the accretion disk has been dissipated, a magnetized stellar-wind torque is applied to the convective region. However, no physical description of the star-disk interaction process was included and it was simply assumed that the rotational period remained constant during the disk accretion phases (`disk-locked' state).  

In the present paper, we include the impact of a self-consistent star-disk interaction model, encompassing the effect of MEs and APSWs on the early PMS stellar angular momentum evolution. In the following sections, we present the main features of our angular moment evolution model.

\subsection{Rotational evolution}
The temporal evolution of the surface stellar angular velocity is governed by the angular momentum evolution of the stellar convective envelope :
\begin{eqnarray}
\rm \dot{J}_{conv} = I_{conv} \dot{\Omega}_\star + \dot{I}_{conv}\Omega_\star = \Gamma_{ext} + \Gamma_{c-e} + \Gamma_{rad} \; ,
\label{jevol}
\end{eqnarray}
where $\rm J_{conv} = I_{conv}\Omega_\star$ is the angular momentum of the convective envelope, $\rm I_{conv}$ is its moment of inertia, and $\rm \Omega_\star$ its angular velocity. The term $\rm \Gamma_{ext}$ is the sum of all of the external torques acting on the stellar surface, while $\rm \Gamma_{c-e}$ is the torque associated with the core-envelope angular momentum exchange and $\rm \Gamma_{rad}$ is the angular momentum variation of the convective envelope due to the development of the radiative core \citep[see e.g.][]{Allain98,GB15,Amard16}. We note that $\rm \dot{I}_{conv}\Omega_\star$ is related to the variation of angular velocity due to the change of the moment of inertia of the convective envelope. This takes into account the stellar contraction and the growth of the core mass, whereas we have neglected the change of the moment of inertia due to mass accretion from the surrounding disk and mass loss associated with stellar outflows. In the following we refer to `internal torque' as $\rm \Gamma_{int} = \Gamma_{c-e} + \Gamma_{rad}- \dot{I}_{conv}\Omega_\star$. 

To compute the torque $\rm \Gamma_{c-e}$, we follow \citet{McGB91}. It implies that both the core and the envelope are in solid body rotation with different angular velocities. A quantity $\rm \Delta J$ of angular momentum is then exchanged between the core and the envelope over a time-scale $\rm \tau_{c-e}$ (hereafter the core-envelope coupling time-scale). The quantity $\rm \Delta J$ is the amount of angular momentum that the core and the envelope {have to} exchange instantaneously in order to have the same angular velocity. We also assume, as in \citet{Allain98}, that $\rm \tau_{c-e}$ is constant for a given model.

The `disk-locked' condition assumed in \citet{Gallet13,GB15} implies that during the accretion phase $\rm \dot{\Omega}_\star = (\rm \Gamma_{ext}+\rm \Gamma_{int})/I_{conv} = 0$, which gives $\rm \Gamma_{ext} =  -\Gamma_{int}$. Here we remove this simplifying assumption by taking into account the internal torques and providing an explicit expression for the external torques acting during the disk lifetime. These terms depend on the characteristics and the structure of the star, summarized in Section \ref{stellarparam}, and the torques associated with the magnetospheric star-disk interaction, presented in Section \ref{SDI}.

\subsection{Stellar parameters, {internal rotation}, and magnetic field}
\label{stellarparam}

To follow the evolution of the stellar structure from the PMS to the end of the MS, we used the standard solar mass model at solar metallicity $\rm Z=0.013446$ \citep{AGSS09}, from the grid of \citet{Amard19} computed with STAREVOL, with a mixing length parameter $\rm \alpha = 1.973$ and an initial helium mass fraction Y, which is equal to 0.269. {The initial time $\rm t_0$ of the model is 970 yrs but we only start to display the evolution at 1 Myr since there are no observations before the Orion Nebula Cluster (hereafter ONC) age.} It is important to note that the structure models do not include the effects of accretion or rotation, and they evolve at a constant mass during the disk-coupling phase.

All of the external torques that are presented in Section \ref{SDI} require the intensity of the dipole component of the stellar magnetic field $\rm B_{dip}$ as input. In \cite{Gallet13, GB15}, $\rm B_{dip}$ was initially identified as the mean magnetic field $\rm B_{CR}$ estimated by the BOREAS\footnote{\url{http://lasp.colorado.edu/~cranmer/Data/Mdot2011/}} routine \citep{Cranmer11} that provides the temporal evolution of the mean magnetic field intensity as a function of stellar parameters. In \citet[][]{Cranmer11} the mean magnetic field $\rm B_{CR}$ is given by
\begin{eqnarray}
\rm B_{CR} &=& 1.13 \rm \sqrt{\frac{\rm 8 \pi \rho_* k_B T_{eff}}{\rm \mu m_H}} \frac{0.55}{\left[1 + (x/0.16)^{2.3}\right]^{1.22}},
 \label{bf}
\end{eqnarray}
where $\rm x = Ro / Ro_{\odot}$ with $\rm Ro = P_{rot}/\tau_{conv}$ the Rossby number with $\rm P_{rot}$ the rotation period, $\rm \tau_{conv}$ the convective turnover timescale \citep[see][for more details about the convective turnover timescale]{Wright11,Oglethorpe13}, $\rm Ro_{\odot}=1.96$, $\rm \rho_*$ the photospheric mass density, $\rm k_B$ the Boltzmann's constant, $\rm T_{eff}$ the effective temperature, $\mu$ the mean atomic weight, and $\rm m_H$ the mass of a hydrogen atom.

\begin{figure}
\includegraphics[angle=-90, width=\linewidth]{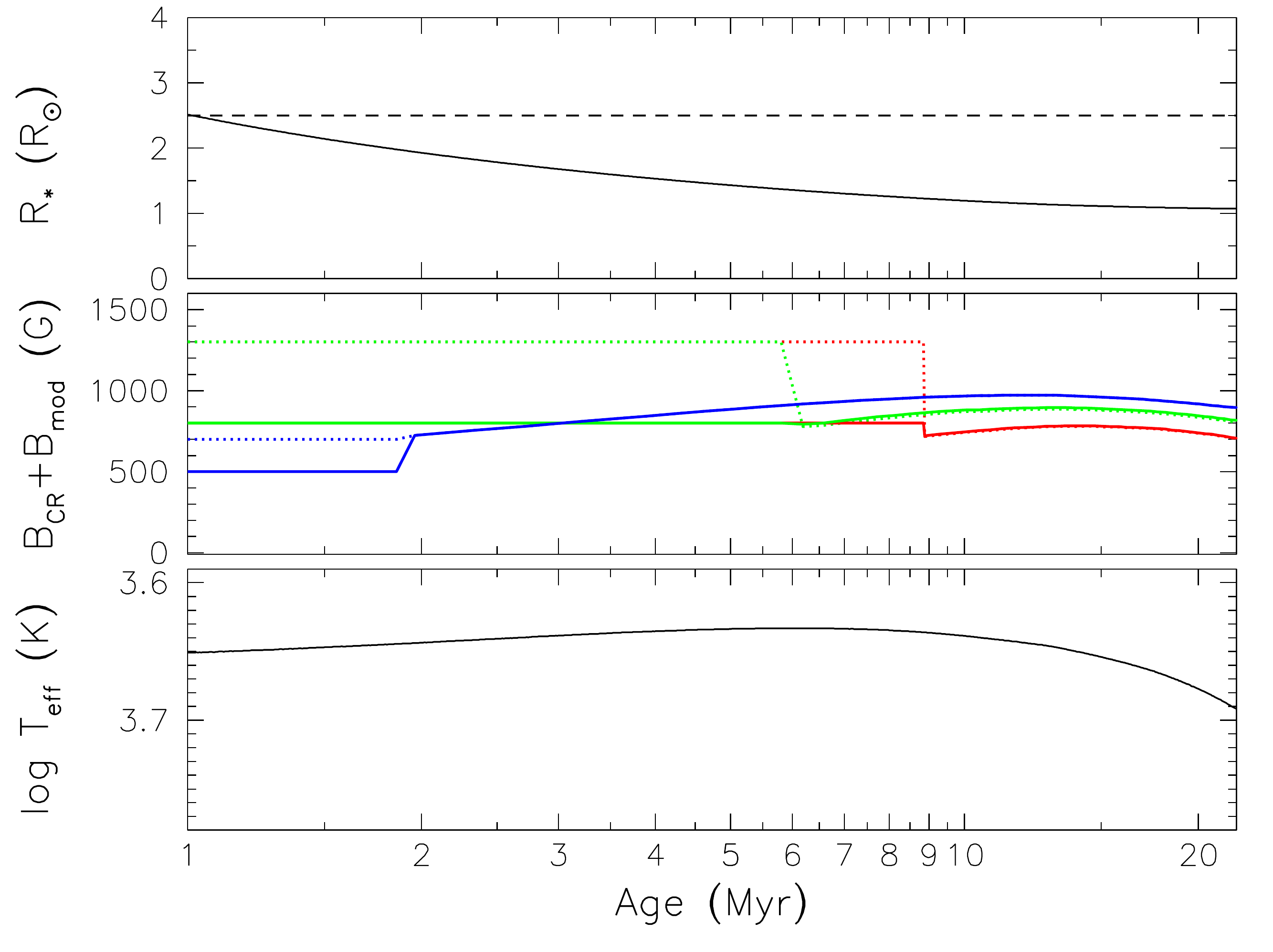}
\caption{Evolution of stellar radius $\rm R_{\star}$ $\rm (R_{\odot})$ (upper panel), magnetic field $\rm B_{CR}$ + $\rm B_{mod}$ (G) (middle panel), and effective temperature $\rm T_{eff}$ (K) (lower panel) as function of time. It is for a 1 $\rm M_{\odot}$ star at solar metallicity and for accretion rate at 1 Myr of $\rm \dot{M}_{acc,init}=10^{-9} M_{\odot}/yr$. The stellar model is from \citet{Amard19}. The dashed line in the upper panel corresponds to $\rm R_{\star} = 2.5~R_{\odot}$. In the middle panel, the colours correspond to the three different initial rotation rates: fast (blue), median (green), and slow (red). The solid lines are for $\rm Q_{acc} = 10\%$ and the dotted lines are for $\rm Q_{acc} = 1\%$.}
\label{evolparam}
\end{figure}

The BOREAS routine reproduces reasonably well the mean magnetic field and the mass loss rate of the present day Sun ($\langle B_{\odot} \rangle\approx$ 2-7 G and $\rm \dot{M}_{\odot} \approx 2~10^{-14} M_{\odot}.yr^{-1}$). We note that for less massive stars, this formalism produces magnetic fields with smaller intensity than what is measured using spectropolarimetry \citep{Morin08,Morin10}. When extrapolated to convective TTauri stars, this model does not provide magnetic fields consistent with the total mean stellar field intensity, but the strength produced by the BOREAS model is compatible with the dipolar components of CTTS (a few hundred Gauss\footnote{It is important to note that in our models, the $\rm B_{dip}$ value corresponds to the dipolar field intensity at the magnetic equator, which is the value entering the definition of the magnetic dipole moment. Spectropolarimetric observations that perform spherical harmonics decomposition of the surface stellar field, usually provide the values of the components at their magnetic pole \citep[see e.g. ][]{Donati2011}, which in the case of a dipole is twice its equatorial strength.}). We assume that our torque models only depend on the dipolar field intensity since, due to its slower radial decrease, the dipolar component is mainly responsible for the large-scale magnetic interactions of the star with its surroundings. 

To further investigate the magnetic field strength required by the star-disk interaction processes to prevent the star from spinning up, we introduced a dipolar magnetic field intensity $\rm B_{\rm mod}$ that we assume is constant during the disk lifetime. This magnetic field strength was chosen so as to best reproduce the rotational distribution observed in the early-PMS phase. After the disk dissipation, the numerical model switches to the mean magnetic field $\rm B_{CR}$, but the transition between $\rm B_{mod}$ and $\rm B_{CR}$ can be very abrupt and thus not physically consistent. 

{Figure \ref{evolparam} shows the evolution of the stellar radius ($\rm R_{\star}$), some examples of the magnetic field intensity ($\rm B_{CR}$ + $\rm B_{mod}$), and the effective temperature ($\rm T_{eff}$) as a function of time. It is important to notice that the magnetic field evolution changes with the parameters that define our models, see Section \ref{results}. In this figure, the transition between a disk and disk-less regime is depicted by the sharp decrease of the magnetic field intensity. It corresponds to the shift between the constant required magnetic field $\rm B_{mod}$ to the BOREAS' magnetic field $\rm B_{CR}$.}

\subsection{Star-disk interaction and associated torques}
\label{SDI}
The star-disk interaction is described in the framework of the scenario proposed by \citet{Zanni2013}. The model is based on axisymmetric magnetohydrodynamic numerical simulations of the interaction of a purely dipolar magnetosphere with the surrounding accretion disk.
In this context the star can exchange angular momentum with its surrounding environment in three different ways: firstly, through the accretion of angular momentum from the disk; secondly, via the action of {MEs} associated with the inflation, expansion, and reconnection process of closed magnetic field lines connecting the star and the disk; and thirdly with angular momentum loss due to stellar winds. Indeed, once the disk has dissipated, stellar winds remain the only spin-down torque. 
%
%

\subsubsection{Accretion}
\label{acc}

According to \citet{Zanni2013}, if the stellar magnetic field is strongly coupled to the accretion disk (i.e. assuming that the disk material is characterized by a low enough magnetic resistivity), the stellar magnetosphere is steadily connected with the disk over a limited radial extent around the truncation radius $\rm R_t$. At this radius, the accretion flow is deviated to form the accretion curtains and the star directly accretes mass and angular momentum from the disk.

We assume that the mass accretion rate evolves according to a simple decay function based on \citet[][i.e. $\rm \dot{M}_{acc} \propto t^{-1.2}$]{Caratti12}
\begin{eqnarray}
\rm f_{decay}(t) = \left(\frac{t_{disk}}{t_{init}}-1\right)^{-1.2}  \left(\frac{t_{disk}}{t}-1\right)^{1.2},
\label{fdecay}
\end{eqnarray}
where $\rm t_{init}$ is the starting age of our simulation (i.e. $10^6$ yr), $\rm t_{disk}$ the disk lifetime, and $\rm t$ the age considered. The mass accretion rate is thus defined as $\rm \dot{M}_{acc} (t)= \dot{M}_{acc,init} f_{decay}(t)$.

We parametrize the accretion torque $\rm \Gamma_{acc}$ as 
\begin{eqnarray}
\rm \Gamma_{acc} = K_{acc}\dot{M}_{acc} \sqrt{\rm GM_\star R_t},
\label{tacc}
\end{eqnarray}
which is proportional to the mass accretion rate $\rm \dot{M}_{acc}$ and to the disk specific Keplerian angular momentum in the truncation region. As it is better specified in the next section, the proportionality constant $\rm K_{acc}$ takes into account the fact that in the truncation region, the disk is not in Keplerian rotation because of disk outflows that have extracted a relevant fraction of the disk angular momentum. The radius $\rm R_t$ corresponds to the region around the star where the dipolar magnetosphere truncates the accretion disk. It can be approximated by \citep{Bessolaz08}:
\begin{eqnarray}
\rm R_t = K_t \left(  \frac{B_{dip}^4R_\star^{12}}{GM_\star\dot{M}_{acc}^2} \right)^{1/7},
\label{rt}
\end{eqnarray}
where $\rm K_t $ is a {dimensionless parameter}. Following \citet{Long05} and \citet{Zanni09}, we assume $\rm K_t=0.5$.

\subsubsection{Magnetospheric ejections}
\label{ME}
{The MEs} \citep[][]{Zanni2013} result from the interaction of a stellar magnetosphere and an accretion disk that produces the expansion and reconnection of the magnetic field lines connecting the star with its surrounding disk. The resulting inflation at mid-latitudes of the dipolar magnetic field lines is very dynamic and goes along with outflows that {can} exchange mass and angular momentum with both the disk and the star. Because of the magnetic field lines reconnection, these outflows detach from the magnetosphere and continue their propagation as magnetic islands, which are disconnected from the central part of the star-disk system and are in between the open magnetic surfaces that are anchored into the star and those anchored into the disk \citep[see Fig. 2 from][]{Zanni2013}.

The opening of the magnetospheric field lines limits the size of the area over which the star and the disk are magnetically connected, which typically does not extend beyond the corotation radius:
\begin{eqnarray}
\rm R_{co} = \left( \frac{G M_\star}{\Omega_\star^2} \right)^{1/3}.
\label{rco}
\end{eqnarray}
Beyond this radius, the disk and the star do not have a direct magnetic connection. Therefore, the \cite{GhoshLamb1979} scenario cannot be directly applied: the disk region beyond corotation, which rotates slower than the star, cannot exert any direct spin-down torque onto the star. 
{Here, we make the assumption that the large-scale magnetic field is responsible for the angular momentum exchange between the different parts of the system. With this approximation, angular momentum and electric currents flow along magnetic field lines. The angular momentum is deposited, or extracted, where the electric current closes perpendicularly to the field, inducing a net Lorentz force.
In \citet[][see Section 4]{Zanni09} it is shown that in order to have an angular momentum exchange between the star and the disk region beyond the corotation radius, a magnetic connection between the two is needed. The star-disk differential rotation generates an electric current flow that closes inside the disk along these field lines, thus depositing a fraction of the stellar angular momentum.
We note that the solutions presented in \citet[][see Appendix A]{Zanni2013}, that are used here to constrain the torque models, do not display such a large-scale star-disk magnetic connection. In these simulations, only angular momentum exchanges between the star and the stellar wind, the MEs, and the part of the disk below corotation are possible.}

As it is shown in \citet{Zanni2013}, one effect of the MEs is to reduce the `Keplerian' accretion torque. This is qualitatively expressed by the constant $\rm K_{acc} < 1$ that translates the fact that a fraction $\rm 1-K_{acc}$ of the accretion torque is launched in the form of MEs. This effect is similar to the action of an X-Wind, which represents the limiting case with $\rm K_{acc} = 0$ for which all the disk angular momentum is extracted by the wind and the disk exerts no torque onto the star. In the case of MEs, we assume a lower efficiency and employ $\rm K_{acc} =  0.4$.

The torque directly exerted by the MEs onto the star is related to a differential rotation effect between the star and the MEs. The torque exerted by stellar magnetic field lines coupled to a region of size $\rm \Delta R$ of the MEs can be expressed as \citep{AC96,MP05a}
\begin{eqnarray}
\rm \Gamma_{ME} = q \Delta R \frac{B_{dip}^2 R_\star^6}{R_t^4},
\label{tdown}
\end{eqnarray} 
where $\rm q = B_{\rm \phi} / B_z \propto \left[ \Omega_{MEs} - \Omega_\star\right] / \Omega_{MEs} $ \citep{LP92,AC96,MP05b} takes into account the differential rotation between the star and the MEs. This factor can be approximated as:
\begin{eqnarray}
\rm q \propto K_{rot} - \left( \frac{R_t}{R_{co}} \right)^{3/2},
\label{qzanni}
\end{eqnarray}
with $\rm K_{rot} = 0.7$. A $\rm K_{rot} < 1$ expresses the fact that the MEs rotate at a sub-Keplerian rate. Assuming $\rm \Delta R \propto R_t$, the MEs torque can be written as:
\begin{eqnarray}
\rm \Gamma_{ME} = K_{ME} \frac{B_{dip}^2R_\star^6}{R_t^3} \left[ K_{rot} -  \left(\frac{R_t}{R_{co}} \right)^{3/2} \right],
\label{tme}
\end{eqnarray}
where $\rm K_{ME} = 0.21$. It is interesting to note that this torque spins-down the star only for $\rm R_t > K_{rot}^{2/3}R_{co}$ while for a lower truncation radius, the MEs spin the stellar surface up.

We used the results of the simulations presented in \citet{Zanni2013} to calibrate the constants appearing in Eqs. \ref{tacc} and \ref{tme}. These results explore only a small fraction of the parameter space of the star-disk interaction problem and even this limited sample suggests that the $\rm K_{acc}, K_{rot},and    K_{ME}$ values are not constant and are a function of the parameters of the model, such as the mass accretion rate. However, the simple formulation adopted in this paper allows one to reproduce the main feature of the MEs scenario, such as the decrease of the accretion torque or the spin-up and spin-down change depending on the relative position of the truncation and corotation radii.

\subsubsection{Stellar winds}
\label{stellawind}

The torque exerted by the magnetized stellar winds on the stellar surface can be expressed as \citep[see][]{WD67}
\begin{eqnarray}
\label{djdt}
\rm \Gamma_{wind} = \dot{M}_{wind} \; r_A^2 \; \Omega_\star
\end{eqnarray}
where  $\rm \dot{M}_{wind}$ is the wind mass loss rate and $\rm r_A$ is the average value of the Alfv\'en radius. We use the expression obtained by \citet{Matt12}
\begin{eqnarray}
\label{ranew}
\rm r_A = K_1 \left[ \frac{B_p^2 R_\star^2}{\dot{M}_{wind} \sqrt{K_2^2v_{esc}^2 + \Omega_\star^2 R_\star^2}}\right]^m R_\star,
\end{eqnarray}
where $\rm v_{esc}=\sqrt{2GM_\star/R_\star}$ is the escape velocity. 
We assume the values $\rm m=0.2177$ and $\rm K_2=0.0506,$ provided by \citet{Matt12}. {The constant $K_1$ is given in Table~\ref{param}.}

In agreement with the APSW scenario \citep{MP05b}, we assume that during the disk-accretion phase, the stellar wind takes its driving power from a fraction of the energy dissipated by the impact of the accretion columns onto the stellar surface. This establishes a relation between the accretion and the wind driving power and consequently between the mass accretion and the wind outflow rate. We take into account two values for the accretion ratio $\rm \dot{M}_{wind}/\dot{M}_{acc} = Q_{acc} = 1\%$ and $10\%$\footnote{This assumption is equivalent to assume a $\rm K_1 = 1.3$ constant, as originally found by  \citet{Matt12}, and $\rm Q_{acc} \approx 3\%$ and $\rm Q_{acc} \approx 30\%,$ respectively.}.
 
\begin{figure}
\includegraphics[angle=-90, width=\linewidth]{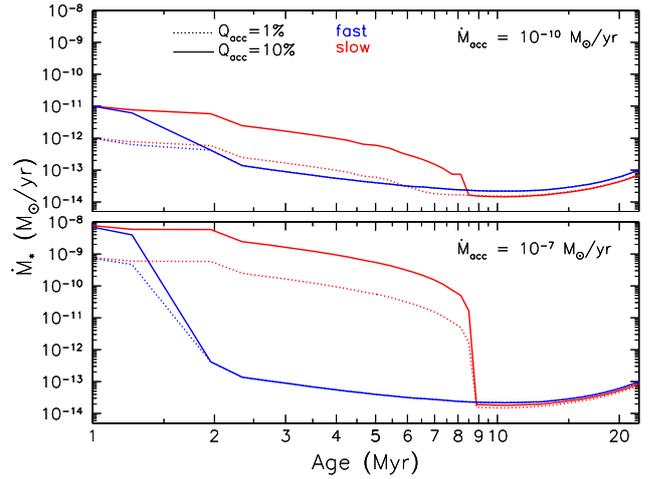}
\caption{Evolution of $\rm \dot{M}_{\star}$ as function of time in case of fast (blue) and slow (red) initially rotating stars. For each rotators, two initial mass accretion rates are investigated $\rm \dot{M}_{acc}=10^{-10}$ (upper panel) and $10^{-7}$ (lower panel) $\rm M_{\odot}/yr$ with $\rm Q_{acc} = 1$ (dotted) and $10\%$ (solid).}
\label{mdotevol}
\end{figure}

After the disk has dissipated ($\rm t > \tau_{disk}$), the stellar wind can not derive its power from accretion anymore and the stellar wind torque becomes the only spin-down mechanism left. As in \cite{Gallet13,GB15}, during this phase the mass-loss is estimated using the BOREAS routine \citep{Cranmer11} that uses the stellar angular velocity, luminosity, and radius as input. 
{In order to prevent a sharp transition at $\rm t = \tau_{disk}$ during the star-disk interaction regime, we use the wind mass loss rate as the maximum value between the accretion-powered and the BOREAS's rates.} {Figure \ref{mdotevol} displays the evolution of the mass loss rate as a function of time from the disk accretion phase up to the {late PMS} for some of our models. The figure clearly shows the transition between the APSW mass loss rate and the BOREAS's mass loss rate at $\rm t = \tau_{disk}$.}

\subsubsection{External torque summary}
\label{externaltorque}

\begin{figure}
\resizebox{\hsize}{!}{\includegraphics*{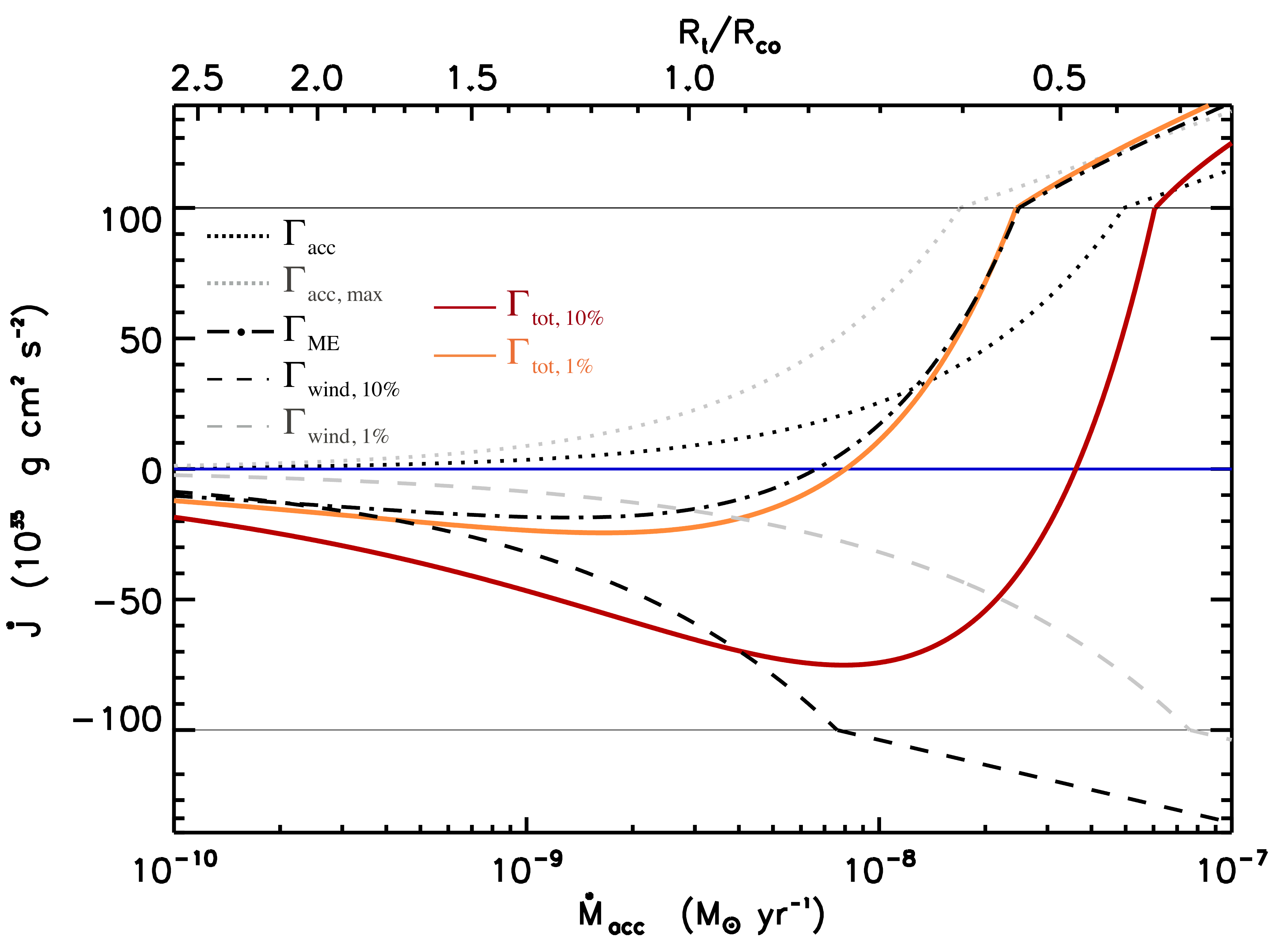}}
\caption{External torques as function of mass accretion rate (lower axis) and corresponding truncation radius (upper axis) for given set of stellar parameters at 1 Myr ($\rm M_\star = 1.0~M_\odot, R_\star = 2.5~R_\odot, P_\star = 7$ days, $\rm B_p = 1$ kG). The accretion torque $\rm \Gamma_{acc}$ ({black dotted line}), the maximum spin-up torque allowed $\rm \Gamma_{acc}~(K_{acc} = 1)$ ({grey dotted line}), the MEs torque $\rm \Gamma_{ME}$ ({dot-dashed line}), the stellar wind torque $\rm \Gamma_{wind}$ for $\rm Q_{acc} = 1\%$ ({grey dashed line}) and $\rm Q_{acc} = 10\%$ ({black dashed line}), and the total external torque $\rm \Gamma_{ext} = \Gamma_{acc} + \Gamma_{ME} + \Gamma_{wind}$ for $\rm Q_{acc} = 1\%$ ({orange solid line}) and $\rm Q_{acc} = 10\%$ ({red solid line}) are plotted.}
\label{modelclaudio}
\end{figure}

In order to point out some important properties of the total external torque $\rm \Gamma_{ext} = \Gamma_{acc} + \Gamma_{ME} + \Gamma_{wind}$ that acts during the accretion phases, we plot in Fig. \ref{modelclaudio} the torques presented in Sections \ref{acc}, \ref{ME}, and \ref{stellawind} as a function of the mass accretion rate for a solar mass star with a radius $\rm R_\star = 2.5~R_\odot$, a rotation period of seven days (i.e. $\approx 0.07$ of the break-up speed), and a dipolar field $\rm B_p = 1$ kG. The black  and grey dotted lines are the accretion torque and the maximum accretion torque (i.e. $\rm K_{acc} = 1$), respectively. The torque associated to the MEs is shown with a  black dot-dashed line, and the grey and black dashed line represent the stellar wind torque for $\rm Q_{acc} = 1\%$ and $\rm Q_{acc} = 10\%$, respectively. Finally, the orange and red solid lines show the evolution of the total external torque for $\rm Q_{acc} = 1\%$ and $\rm Q_{acc} = 10\%$, respectively.

As the disk truncation gets closer to the stellar surface, $\rm \Gamma_{ext}$ assumes positive values (i.e. it exerts a spin-up torque) and asymptotically approaches the limiting value $\rm \Gamma_{ext} \approx \dot{M}_{acc} \sqrt{GM_\star R_t}$ (i.e. $\rm \Gamma_{acc}$ with $\rm K_{acc} = 1$). This coincides with the fact that, in such a situation, the external torque is dominated by accretion and the spin-down effects associated with MEs become negligible \citep{Zanni2013}. In this regime, even a massive stellar wind ($\rm Q_{acc} = 10\%$) is not able to balance the accretion torque.
One can notice that, in contrast with the \cite{Zanni2013} simulations, in this regime the spin-up torque associated to the MEs is even larger than the spin-up torque due to accretion. This is a consequence of having assumed constant $\rm K_{ME}$ and $\rm K_{acc}$ values, while the \cite{Zanni2013} simulations suggest that $\rm K_{acc}$ ($\rm K_{ME}$) increases (decreases) with the accretion rate and the $\rm R_{t}/R_{co}$ value. On the other hand, the $\rm \Gamma_{acc} + \Gamma_{ME}$ sum becomes comparable to the maximum total spin-up torque ($\rm \Gamma_{acc}$ with $\rm K_{acc} = 1$), which is consistent with the simulations employed to calibrate our models. 

For $\rm Q_{acc} = 1\%$, the spin-down torque due to MEs becomes comparable to the stellar wind's torque for $\rm R_t / R_{co} \approx 1$ and becomes more important for larger $\rm R_t / R_{co}$ values. For $\rm Q_{acc} = 10\%,$ the MEs and stellar wind spin-down torques become comparable for even larger $\rm R_t / R_{co}$ values, corresponding to a strong propeller regime \citep{IS75} in which the stellar centrifugal barrier prevents the disk from accreting (see Section \ref{interregime}). For $\rm Q_{acc} = 1\%$, and even more so for $\rm Q_{acc} = 10\%$, in a steadily accreting regime (i.e. $\rm R_t < R_{co}$), the spin-down torque that is due to stellar winds is always more important than the MEs' torque.

In agreement with the \citet{Zanni2013} results, the $\rm \Gamma_{ext} = 0$ condition (i.e. when the spin-down torques associated with the MEs and the stellar wind exactly balance the accretion torque) is already attained for $\rm R_t < R_{co}$ even in the presence of a weak stellar wind ($\rm Q_{acc} = 1\%$). This condition is clearly not sufficient to keep the stellar angular velocity constant, which requires the spin-down torques to balance both accretion and internal torques ($\rm \Gamma_{int} \neq 0$). For both values of $\rm Q_{acc,}$ it is possible to define an optimal configuration, corresponding to the minimum of the $\rm \Gamma_{ext}$ curves, as something that is related to the most efficient spin-down torque that can be obtained for a given set of stellar parameters. For $\rm Q_{acc} = 10\%,$ such a configuration is obtained for $\rm R_t \leq R_{co}$ and the spin-down torque is dominated by the stellar wind. For $\rm Q_{acc} = 1\%$, the maximum spin-down is attained for $\rm R_t \geq R_{co}$ and the MEs spin-down torque becomes more important than the stellar wind torque. This is consistent with the \citet{Zanni2013} results, which show that the system must enter a propeller regime in order for the star-disk interaction's spin-down torque to be strong enough to balance the stellar contraction.

As the truncation radius moves further out due to the decrease of the accretion rate, the total external torque still leads to a spin-down, but it gets weaker. For a given value of the stellar magnetic field, the increase of the differential rotation between the star and the MEs in Eq. \ref{tme} does not balance the weakening of the magnetic field intensity in the truncation region, while the stellar wind torque weakens since its mass outflow rate decreases.

\section{Results}
\label{results}

In order to compare with the observed distributions of stellar periods of rotation, we computed rotational evolution models by varying the initial (at 1 Myr) period of rotation, accretion rate, dipolar field intensity, and stellar wind mass-loss rate. 

\subsection{Rotational period constraints and models}

{We used the rotation period distribution of five star-forming regions from 1.5 Myr \citep[The Orion Nebula Cluster,][]{RL2009} to 13 Myr \citep[the h PER cluster,][]{Moraux13}} as observational constraints for the angular velocity evolution of solar-like stars. The age and the references for each cluster are listed in Tab. \ref{opencluster}. {Among these clusters, ONC, NGC 6530, Cep OB3B, and NGC 2362 display near-infrared excesses that are associated to the presence of circumstellar disks and by extension to a disk-coupling phase \citep{Bell13}.}

\begin{table}
\caption{Open clusters whose rotational distributions are used is this study.}   
\label{opencluster}      
\centering                          
\begin{tabular}{c c c}       
\hline\hline                
Cluster & Age & Ref. \\  
 &(Myr)&  \\
\hline   
ONC & 1.5& 1  \\
NGC 6530 & 2 & 2 \\
Cep OB3b & 4 & 3 \\
NGC 2362 & 5  & 4 \\
h PER & 13  & 5 \\
\hline   
\end{tabular}
\tablebib{(1)~\citet{RL2009}; (2) \citet{Henderson11}; 
 (3) \citet{Littlefair10}; (4) \citet{Irwin08a}; (5) \citet{Moraux13}.} 
\end{table}

We computed models for slow, median, and fast rotators, which are characterized by a different initial period of rotation at 1 Myr. The initial rotation periods were chosen following \citet{GB15} so as to be able to compare the rotational tracks of this paper with the ones from their work.
{The different types of rotators are} characterized by a core-envelope coupling time-scale $\rm \tau_{c-e}$, {a} disk lifetime $\rm \tau_{disk}$, and {an }initial rotational period $\rm P_{init}$ (see Table \ref{param}). We computed the evolution of each rotator for four different initial mass accretion rates in the range $\rm \dot{M}_{init} = 10^{-10} - 10^{-7} \, M_\odot$.yr$^{-1}$ and used two possible values of the stellar wind mass loss and accretion rate ratio during the accretion phases, $\rm Q_{acc} = 1\%$ and $10\%$. After the disk dissipates, the wind mass loss rate that is provided by the BOREAS model \citep{Cranmer11} is employed, as in \cite{Gallet13,GB15}, see Fig. \ref{mdotevol}. 

During the disk accretion phases, the following two prescriptions were adopted for the dipolar field {strength}: firstly, $\rm B_{CR}$ (the \citet{Cranmer11} magnetic field) as provided by the BOREAS model; and secondly, $\rm B_{mod}$, a constant value chosen to better reproduce the observed rotational distributions. The $\rm B_{mod}$ value was chosen so that the rotational tracks fit the initial stellar period at 1 Myr and the corresponding percentile of the h PER cluster at 13 Myrs. {Clusters older than the h PER cluster are thus not used to constrain $\rm B_{mod}$.} 
{It is important to notice that we do not try to reproduce a perfectly constant spin evolution, but we follow the general assumption that a star in a given rotation state (fast/median/slow) remains in that regime from the early-PMS up to the early-MS phase. The underlying hypothesis is that the initial conditions dictate the evolution of the surface rotation rate of the stars during these phases.}
We recall that a constant $\rm B_{mod}$ is only {used} during the star-disk interaction phase and in both cases the \citet{Cranmer11} magnetic field is used to follow the angular velocity evolution after the end of the disk lifetime. The values of $\rm B_{CR}$ at 1 Myr, its maximum value reached during the disk lifetime, and $\rm B_{mod}$ are given in Table \ref{maccprot}. 

\begin{figure*}
\centering
\includegraphics[angle=-90,width=16cm]{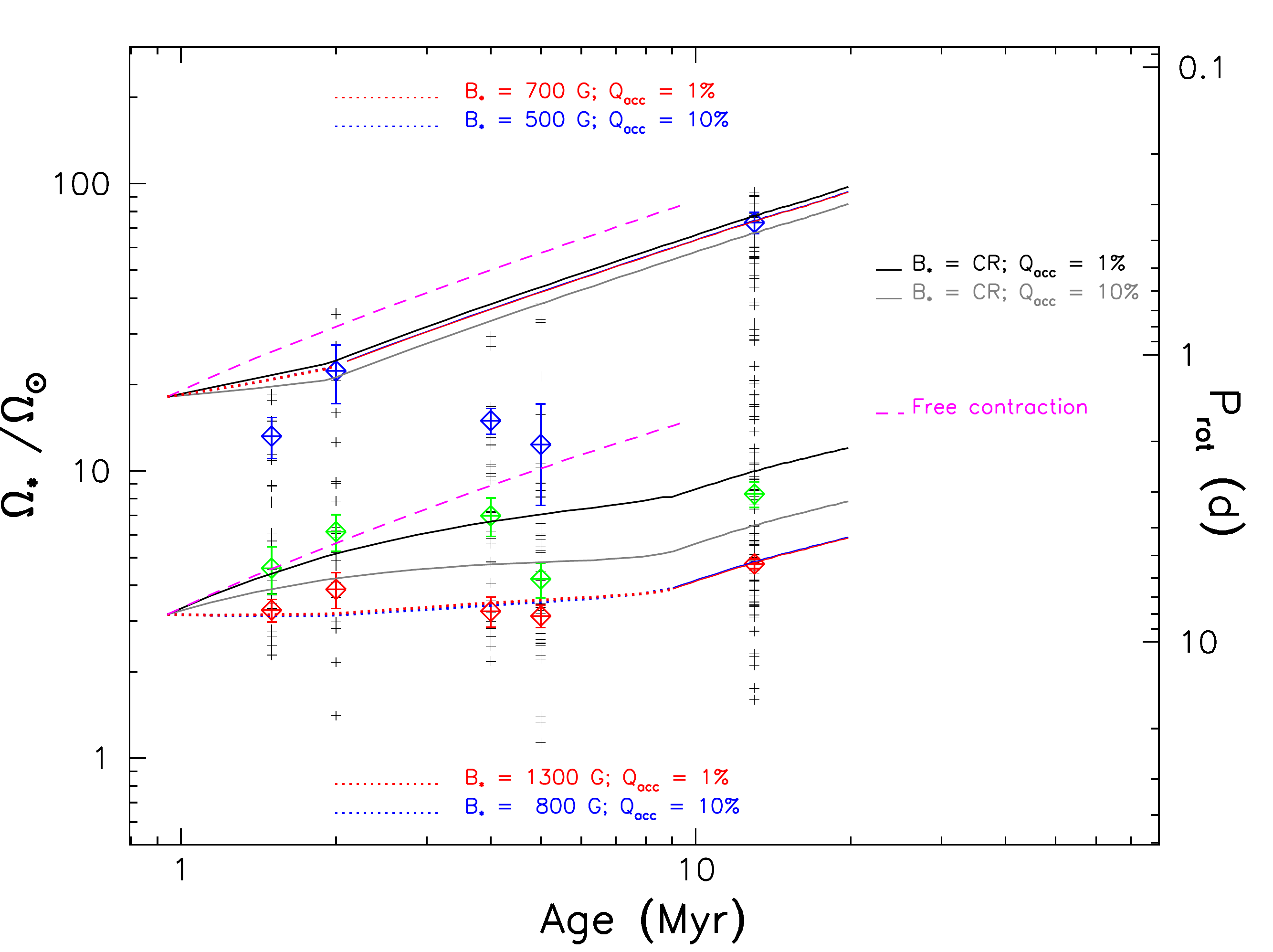}
\caption{Angular velocity evolution of stellar convective envelope as function of time for fast ({up}) and slow ({down}) rotator models in case where $\rm \dot{M}_{acc,init} = 10^{-9}~M_{\odot}.yr^{-1}$. The black and the grey solid lines represent the MEs and APSW processes with $\rm Q_{acc}=1\%$ and $\rm Q_{acc}=10\%$, respectively, and the magnetic field is from the BOREAS routine from \citet{Cranmer11}. The red and the blue dotted lines represent the {models including the} MEs and APSW processes with $\rm Q_{acc}=1\%$ and $\rm Q_{acc}=10\%$, respectively, but with the numerically imposed magnetic field {B$_\textrm{mod}$}. The free contraction (i.e. without any external interaction) is shown by the magenta long-dashed line. It represents the increase of the rotation rate during a free contraction phase. The angular velocity is scaled to the angular velocity of the present Sun. The blue and the red tilted square and associated error bars represent, respectively, the 90th percentile and the 25th percentile of the rotational distributions of solar-type stars in star-forming regions and young open clusters obtained with a rejection sampling method described in \citet[][]{Gallet13}.} 
\label{modelCRBmod}
\end{figure*}

{Figures \ref{modelCRBmod} and \ref{modelCRBmod108} display the surface angular velocity evolution of fast and slow initially rotating stars as a function of age for two different accretion rates $\rm \dot{M}_{acc,init} = 10^{-9}~M_{\odot}.yr^{-1}$ and $\rm 10^{-8}~M_{\odot}.yr^{-1}$, respectively. In these figures, we compare the angular velocity evolution resulting from using the MEs and APSW processes at the two $\rm Q_{acc}$ efficiency, either by using the \citet{Cranmer11} magnetic field $\rm B_{CR}$ (blue and red line) or the imposed magnetic field $\rm B_{mod}$. It is important to note that we only display the first 20 Myr of the evolution because it is at the core of the present work.}

{In these figures, it is clear that for the initially slow rotating stars, the MEs and APSW processes with the imposed magnetic field $\rm B_{mod}$ better match the early-PMS observation than the models using the Cranmer's magnetic field. For these slow rotators, it suggests that a stronger magnetic field, of the order of the kG, is already required at early ages < 1 Myr.}
{However, for the initially fast rotating stars, it seems that both $\rm B_{mod}$ and $\rm B_{CR}$ are quite close to each other. It suggests that the Cranmer's magnetic field is in that case well suited to be used in the framework of the MEs and APSW processes. However, the rotational evolution of these rotators clearly overestimates the 90th percentile of ONC, Cep OB3b, and NGC 2326. This issue was already present in \citet{Gallet13,GB15} and comes from the constraints imposed by the h PER cluster at 13 Myr. Indeed, the contraction rate and the associated increase of the surface rotation due to angular momentum conservation impose a certain initial rotation period that is not compatible with the observed percentile of the rotation distribution in the previously mentioned clusters. We note that it is not an issue linked to the completeness of the rotation period observations; the rotational distribution of Cep OBS3b does indeed seem to be fully recovered for periods $<$ 7 days \citep[see][]{Littlefair10} while the completeness of NGC 2362 is close to 100\% \citep[see][]{Irwin08a}.}
{Additionally, we can see that the free contraction matches the upper part of some of the observations. These stars are more presumably outliers, with initial conditions associated to a very short disk lifetime (< 1 Myr), that therefore populate the fast rotating part of the distributions at ZAMS.}

\begin{table}
\begin{center}
\caption{Model parameters.} 
\label{param}          
\begin{tabular}{c c c c}     
\hline\hline              
Parameter & Slow & Median & Fast \\  
\hline
$\rm P_{init}$ (days) & 8 & 5 & 1.4 \\
$\rm \tau_{c-e}$ (Myr) & 30 & 28 & 10 \\
$\rm \tau_{disk}$ (Myr) & 9 & 6 & 2 \\
$\rm K_1$ & 1.7 & 1.7& 1.7 \\
\hline                                  
\end{tabular}
\end{center}
\end{table}

\begin{threeparttable}
\caption{Magnetic field strength (in G) employed in angular velocity evolution models at 1 Myr. The values for $\rm B_{CR}$, are the maximum magnetic field intensity reached by $\rm B_{CR}$ during the disk lifetime.} 
\label{maccprot}      
\centering
\begin{tabular}[b]{|c|c|c|c||c|c|c|}
 \multicolumn{1}{c}{a) $\rm Q_{acc}= 1\%$} & \multicolumn{3}{c}{$ \rm B_{CR}$} & \multicolumn{3}{c}{$\rm B_{mod}$} \\ 
\hline
$\rm \dot{M}_{acc}$ / $\rm P_{rot}$ (d) & 1.4 & 5 & 8 & 1.4 & 5 & 8 \\
\hline \hline
$\rm 10^{-7}~M_{\odot}.yr^{-1}$ & 653  & 898 & 939 & 500 & 2000 & 2600 \\
$\rm 10^{-8}~M_{\odot}.yr^{-1}$ & 712 & 880 & 914  & 400 & 1200  & 1600 \\
$\rm 10^{-9}~M_{\odot}.yr^{-1}$ & 715  & 882 & 902 & 700 & 1300 & 1300 \\
$\rm 10^{-10}~M_{\odot}.yr^{-1}$ & 655 & 895 & 925  & 1700 & 2400 & 2200 \\ 
\hline
\multicolumn{7}{c}{} \\
 \multicolumn{1}{c}{b) $\rm Q_{acc}= 10\%$} & \multicolumn{3}{c}{$ \rm B_{CR}$} & \multicolumn{3}{c}{$\rm B_{mod}$} \\ 
\hline
$\rm \dot{M}_{acc}$ / $\rm P_{rot}$ (d) & 1.4 & 5 & 8 & 1.4 & 5 & 8 \\
\hline \hline
$\rm 10^{-7}~M_{\odot}.yr^{-1}$  & 640 & 821 & 850  &  300 & 900 & 1300   \\
$\rm 10^{-8}~M_{\odot}.yr^{-1}$  & 699 & 775   & 785 &  300 & 600 & 900    \\
$\rm 10^{-9}~M_{\odot}.yr^{-1}$  & 715 & 839 & 821 &  500 & 800 & 800    \\
$\rm 10^{-10}~M_{\odot}.yr^{-1}$ & 655 & 890 & 911 &  1300 & 1800 & 1600 \\ 
\hline
\end{tabular}
    \begin{tablenotes}
      \small
      \item \textbf{Notes.} The initial values (i.e. at 1 Myr) for $\rm B_{CR}$ are: 590, 530, 440 G for the fast, median, and slow rotators, respectively.
    \end{tablenotes}
\end{threeparttable}

\subsection{Dipolar field intensity}
\label{dipint}

\begin{figure*}
\centering
\includegraphics[angle=-90,width=16cm]{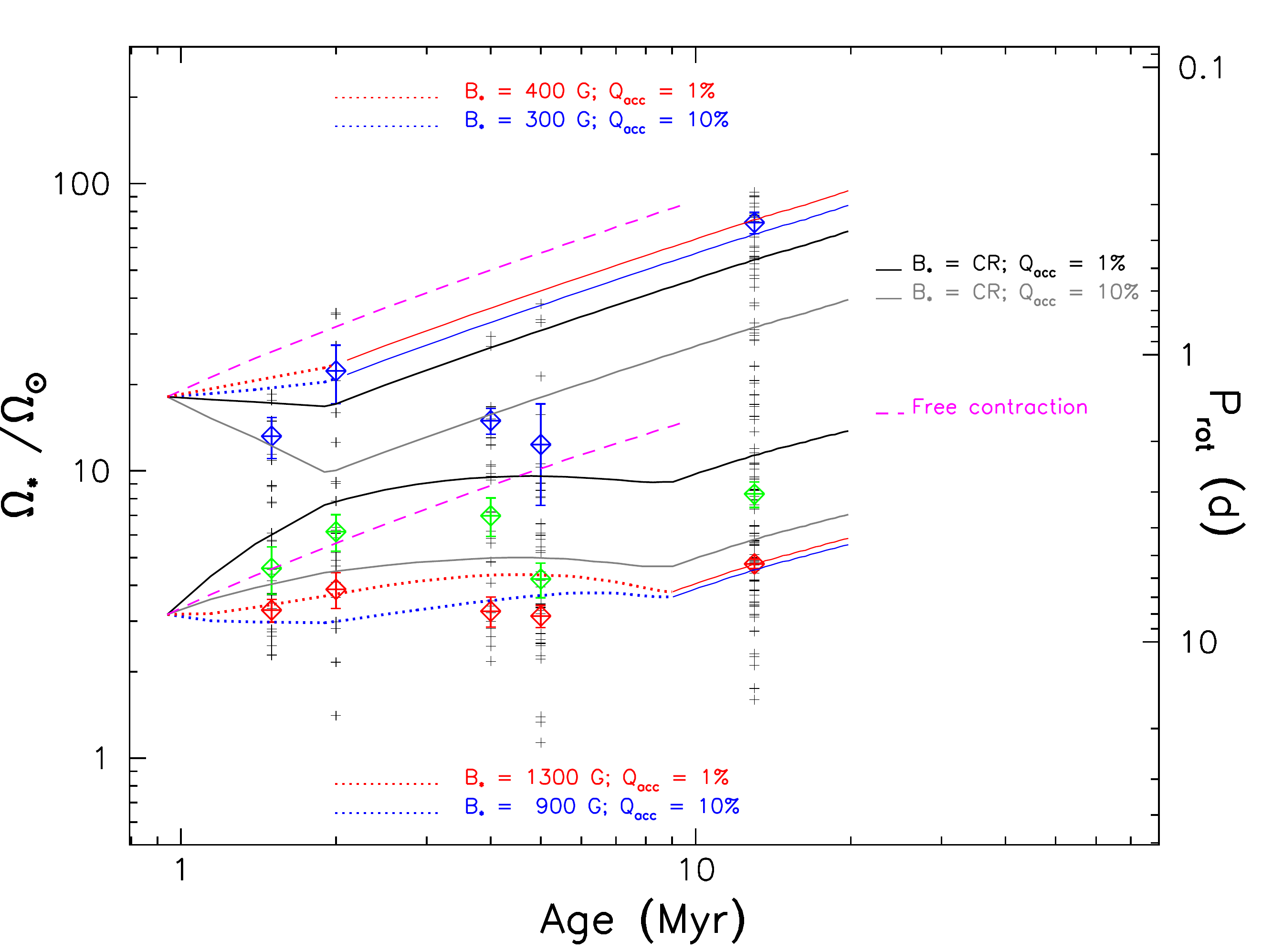}
\caption{Same as Fig. \ref{modelCRBmod}, but for $\rm \dot{M}_{acc,init} = 10^{-8} ~M_{\odot}.yr^{-1}$. It is important to note that for the fast rotation cases, the black and red lines are superimposed.}
\label{modelCRBmod108}
\end{figure*}

In Table \ref{maccprot} we display the dipolar magnetic field values used to compute our models, the $\rm B_{CR}$ value provided by the \citet{Cranmer11} model at 1 Myr, its maximum value during the disk lifetime, and the $\rm B_{mod}$ stay constant during the star-disk interaction phases so as to fit the evolutionary tracks at 1 Myr and 13 Myr. 

We note that $\rm B_{CR}$ is comparable, at least initially, to $\rm B_{mod}$, yielding an approximately constant envelope rotation during the accretion phases for fast rotator models with weaker stellar winds ($\rm Q_{acc} = 1\%$) and accretion rates larger than $\rm 10^{-9}~M_{\odot}.yr^{-1}$. See, for example, the evolutionary track in Fig. \ref{modelCRBmod108} for a fast rotator with $\rm Q_{acc} = 1\%$. 
For fast rotator cases with a larger stellar wind mass-loss rate ($\rm Q_{acc} = 10\%$),, $\rm B_{CR}$ can be even larger than the `optimal' $\rm B_{mod}$ value, corresponding to a rapid spin-down during the star-disk interaction stages, see for example the fast rotator model plotted with a grey line in Fig. \ref{modelCRBmod108}. 

\begin{figure*}
\centering
\includegraphics[width=\linewidth]{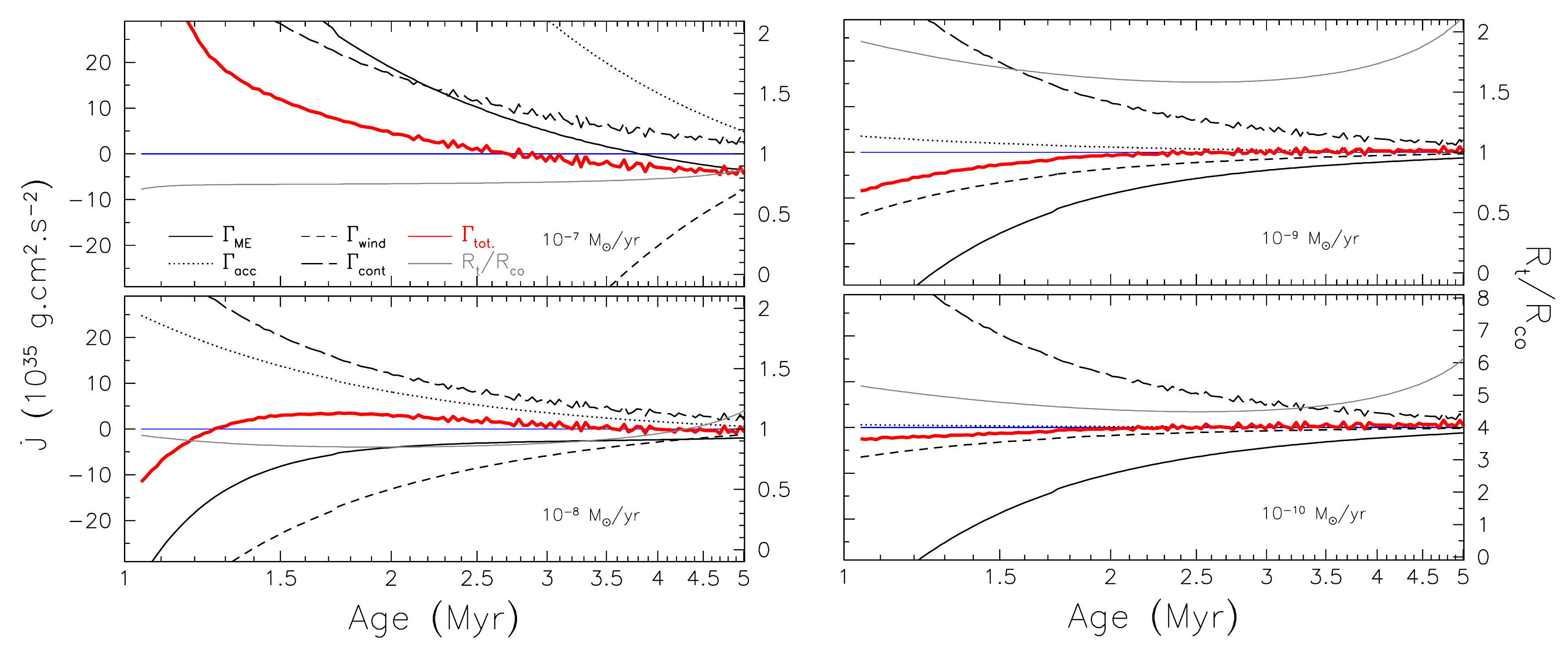}
\caption{Evolution of accretion torque $\rm \Gamma_{acc}$ ({dotted line}),{} ME's torque $\rm \Gamma_{ME}$ ({solid line}), stellar wind {torque} $\rm \Gamma_{wind}$ ({short dashed line}), and{ positive }torque due to stellar contraction ({long dashed line}) as function of time {for} median rotator {with} $\rm Q_{acc} = 1\%$ and $\rm \dot{M}_{acc,init} = 10^{-7}~M_{\odot}.yr^{-1}$ ({upper panel}), $\rm \dot{M}_{acc,init} = 10^{-8}~M_{\odot}.yr^{-1}$ ({upper middle panel}), $\rm \dot{M}_{acc,init} = 10^{-9}~M_{\odot}.yr^{-1}$ ({lower middle panel}), and $\rm \dot{M}_{acc,init} = 10^{-10}~M_{\odot}.yr^{-1}$ ({lower panel}). {The red solid line is the sum of all of these torque contributions.} The blue solid line represents the zero torque condition {leading to }a constant angular velocity during the disk accretion phase. The grey solid line shows the evolution of $\rm R_t / R_{co}$.}
\label{torquemed}
\end{figure*}

In all other cases $\rm B_{mod}$ is larger than the $\rm B_{CR}$ estimate, with values that systematically exceed 500 G (corresponding to 1 kG maximum dipolar intensity at the magnetic pole) up to more than 2 kG (4 kG at the magnetic pole). 
Looking at the behaviour of the dipolar field intensity $\rm B_{mod}$ as a function of the free parameters of the rotational models, it is possible to highlight some trends. Given the same mass accretion rate and rotation period, the $\rm Q_{acc} = 10\%$ models require a weaker field than the $\rm Q_{acc} = 1\%$ cases; a more massive wind clearly provides a more efficient spin-down torque and requires a weaker magnetic field. For a fixed accretion rate, median and slow rotators require a stronger dipolar magnetic component than fast rotator; a more efficient spin-down torque is required to prevent them from spinning-up. For a given rotation period, $\rm B_{mod}$ displays a more complex behaviour as a function of the accretion rate, often passing through a minimum at intermediate accretion rates. This minimum of the dipolar field strength corresponds to the condition of maximum spin-down efficiency characterizing our external torque model outlined in Section \ref{externaltorque} (see Fig. \ref{modelclaudio}). To confirm this result, we plot in Fig. \ref{torquemed} the contribution of the different torques as a function of time in the case of the median rotator model for the four values of $\rm \dot{M}_{acc,init}$ considered and for $\rm Q_{acc} = 1\%$. In the same plots we also show the $\rm R_{t}/R_{co}$ ratio as a function of time. The lower left panel corresponds to the minimum field intensity found in the median rotator column in Table \ref{maccprot}a). As consistent with the torque model discussed in Section \ref{externaltorque}, the star-disk system transits here through a configuration characterized by $\rm R_{t}/R_{co} \approx 1$, roughly corresponding to the minimum (i.e. maximum spin-down efficiency) of the orange solid line in Fig. \ref{modelclaudio}. Similar to the discussion in Section \ref{externaltorque}, for a higher mass accretion rate (a lower $\rm R_{t}/R_{co}$ ratio) the spin-down torque is dominated by the stellar wind. For a lower accretion rate, the MEs provide the main spin-down torque, while, correspondingly, the $\rm R_{t}/R_{co}$ value becomes quite large (up to 5 for $\rm \dot{M}_{\rm acc,ini} = 10^{-10}~M_{\odot}.yr^{-1}$).\ This indicates that the system is most likely in a (strong) propeller regime.
Figure \ref{torquemed} also shows that the total torque is not constant and null in time, which would correspond to a perfect constant $\Omega_\star$ condition. This clearly points to the fact that the star-disk system can go through spin-down and spin-up phases during its evolution. 

\section{Discussion}
\label{disc}

In the Section 3, we show how our modelling allows us to put some constraints on the intensity of the dipolar component of the stellar magnetic field necessary to prevent the star from spinning up during the star-disk interaction phases. Here, we discuss these results, their implications for the star-disk interaction regimes that provide an efficient enough spin-down torque, and the observational constraints and biases that could support or refute our findings.  

\subsection{{A magnetic dipole strength issue}}
The results presented in Section \ref{results} show that, particularly in the case of median and slow rotators, dipolar magnetic fields that are stronger than 1 kG at their magnetic pole are required to efficiently reduce the stellar spin-up, independently of the main spin-down mechanism and the initial disk accretion rate. We recall that our models take into account only two specific mechanisms that can influence the stellar rotational evolution, namely stellar winds and magnetospheric ejections. However, the magnetic field values that we found are consistent with the magnetic field strength found by \citet{Matt10} who use a model similar to \cite{GhoshLamb1979}. As already mentioned in Section \ref{intro}, other possible spin-down scenarios, such as X-winds \citep{Shu94} or ReX-winds \citep{FPA00} were not taken into account due to the lack of a self-consistent model or torque parametrization. 

\begin{table*}
\caption{Values of mass accretion rate, dipole magnetic field strength, and stellar radius from \citet{Johnstone14}. The stars are sorted by increasing dipole magnetic field strength.} 
\label{obs}
\centering
\begin{tabular}{c c c c c c c} 
Star & $\rm B_{dip}$ & $\rm M_{\star}$ & $\rm R_{\star}$ & $\rm P_{rot,\star}$ & $\rm \dot{M}_{acc}$ & $\rm {J_\star}/{\Gamma_{ext}} $  \\
         & (kG) & ($\rm M_{\odot}$) & ($\rm R_{\odot}$) & (days) & (\rm $\rm M_{\odot}.yr^{-1}$) & (1\%/10\%; Myr)\\
\hline\hline
V2247 Oph        & 0.11      & 0.36 & 2.00 & 3.5  & 1.6$\times 10^{-10}$ & -503.2/-25.5 \\
CV Cha           & 0.14      & 2.0  & 2.50 & 4.4  & 3.2$\times 10^{-8}$ & 0.6/0.7 \\
CR Cha           & 0.22      & 1.9  & 2.50 & 2.3  & 1.0$\times 10^{-9}$ & 68.6/-53.4 \\
V2129 Oph 2005   & 0.28      & 1.35 & 2.00 & 6.53 & 6.3$\times 10^{-10}$ & 8.1/13.7 \\
TW Hya 2008      & 0.37      & 0.8  & 1.10 & 3.56 & 1.3$\times 10^{-9}$ & 2.4/3.1 \\
TW Hya 2010      & 0.73      & 0.8  & 1.10 & 3.56 & 1.3$\times 10^{-9}$ & 2.3/3.8 \\
GQ Lup 2011      & 0.90      & 1.05 & 1.70 & 8.4  & 1.0$\times 10^{-9}$ & 2.5/5.3 \\
BP Tau Dec 2006  & 0.96      & 0.7  & 1.95 & 7.6  & 2.5$\times 10^{-9}$ & 1.6/-14.3 \\
V2129 Oph 2009   & 0.97      & 1.35 & 2.00 & 6.53 & 6.3$\times 10^{-10}$ & 17.9/-10.9 \\
GQ Lup 2009      & 1.07      & 1.05 & 1.70 & 8.4  & 1.0$\times 10^{-9}$ & 2.6/7.2 \\
BP Tau Feb 2006  & 1.22      & 0.7  & 1.95 & 7.6  & 2.5$\times 10^{-9}$ & 2.1/-2.8 \\
AA Tau           & 1.72      & 0.7  & 2.00 & 8.2  & 6.3$\times 10^{-10}$ & -3.5/-1.2 \\
\hline
\end{tabular}
\end{table*}

Classical TTauri stars are known to have very strong surface average magnetic fields up to a few kG \citep[][]{JK2007}. Spectropolarimetric observations using the Zeeman–Doppler Imaging (ZDI) technique can also provide constraints on the topology of the stellar field, and they often indicate the presence of a complex field where the dipolar component is not always dominant \citep[see e.g.][]{Johnstone14}.
These reconstructions need a dense and regular sample of rotation period and cold stars. 
One {strong limitation} of this {technique} is that depending on the structure of the magnetic field, some components can cancel each other out \citep{Morin10}, reducing the strength of the magnetic components. Moreover, the reconstructions are done at a specific time $\rm t$, neglecting the possible temporal evolution of the magnetic topology. 

Keeping this in mind, we tried to apply our simple torque model presented in Section \ref{model} to the sample presented in \cite{Johnstone14}, and summarized here in Table \ref{obs}. The dipolar field value corresponds to the intensity at the magnetic pole. 
We also provide an estimate for the characteristic spin-up (positive) or spin-down (negative) timescale. We computed it as the ratio of the total (core plus envelope) angular momentum of the star J$_\star$ and $\Gamma_\textrm{ext}$ obtained with our torque model (see Section \ref{model}), using the stellar parameters displayed in Table \ref{obs}.

Using the $\rm Q_{acc} = 10\%$ model, only AA Tau and BP Tau ($\rm B_{dip} = 1220 \, G$) are characterized by a spin-down timescale of a few million years, which is compatible with the Kelvin-Helmholtz contraction timescale. V2129 Oph ($\rm B_{dip} = 970 \, G$), BP Tau ($\rm B_{dip} = 960 \, G$), CR Cha, and V2247 Oph are in a spin-down state, but the associated timescale is larger than 10 Myrs. GQ Lup, V2129 Oph ($\rm B_{dip} = 280 \, G$), TW Hya, and CV Cha are in a spin-up state. It is possible to notice that the same object can go from a spin-up to a spin-down state at different epochs, as in the case of V2129 Oph, or conversely, the efficiency of the spin-down torque can change, as is the case of BP Tau, which highlights the usefulness of multi-epoch observations. In applying the $\rm Q_{acc} = 1\%$ model, only AA Tau would be characterized by a spin-down timescale shorter than 5 Myrs.

In our models we also made the limiting assumption that the $\rm B_{mod}$ magnetic field, selected to best reproduce the observed rotational evolution, stays constant during the disk accretion phase. Recently \citet{Folsom16,Folsom18} investigated the evolution of the magnetic field strength and topology of low-mass stars from the PMS to the end of MS. They find that up to the ZAMS, the magnetic field properties are primarily driven by the structural evolution of the stars, while during the MS phase the magnetic field strength decreases with a decreasing stellar rotational period.
{Actually, the intensity of the magnetic field rapidly decreases during the first 10 Myr of the stellar evolution \citep{Folsom16} following the increase of the complexity of the internal structure \citep{Gregory12,Villebrun18}.}

\subsection{The interaction regimes}
\label{interregime}
As discussed in Section \ref{dipint}, the $\rm Q_{acc} = 10\%$ models, and according to which the mass outflow rate of the stellar wind during the disk accretion phase corresponds to 10$\%$ of the mass accretion rate, require a weaker dipolar field that  is closer to the typically observed values than the intensities required by the $\rm Q_{acc} = 1\%$ models. 
On the other hand, the feasibility and nature of these powerful stellar winds is still debated. Since TTauri stars are slow rotators that spin at a fraction $\leq 10 \%$ of their break up speed, stellar winds can not be driven by a magneto-centrifugal mechanism. It is unlikely that the driving force is given by a thermal pressure gradient, which requires a coronal temperature close to virial, that is, $\approx 10^6 $ K. \citet{Matt07} show that a massive wind would have an X-Ray luminosity that is much higher than the typical value observed for TTauri stars, and its total radiative power would largely exceed the wind kinetic power and even the accretion luminosity. These estimates can provide an upper limit for a thermally-driven wind $\rm \dot{M}_{wind} \lesssim 10^{-11} \, M_\odot.yr^{-1}$. In order to cool less efficiently, a more massive wind, such as $\rm \dot{M}_{wind} \approx 10^{-9} \, M_\odot.yr^{-1}$, should have a temperature around $10^4$ K, and therefore can not be thermally driven. 
As an alternative, it has been proposed that these winds can be driven by the momentum deposited by Alfv\'en waves \citep{Decampli81}, which are possibly excited and amplified by the impact of the accretion funnels onto the stellar surface \citep[APSW,][]{MP05b}. Moreover, one-dimensional MHD models of this process suggest that the wind mass loss rate can not be higher than 1$\%$ of the mass accretion rate  \citep{Cranmer08}.

On the one hand, if we assume an ejection efficiency of $\rm Q_{acc} = 1 \%$, which is less problematic from the point of view of the stellar wind theory, our models require a stronger dipolar field intensity. Besides, according to our discussion in Sections \ref{externaltorque} and \ref{dipint}, for mass accretion rates lower than approximately $\rm \dot{M}_{acc} < 10^{-8} \, M_\odot.yr^{-1}$, which roughly correspond to the maximum spin-down efficiency and to the minima of the $\rm B_{mod}$ values in Table \ref{maccprot} for the $\rm Q_{acc} = 1\%$ models, the system is likely to enter a `propeller' regime. We recall that a star is in a propeller regime \citep{IS75} when the inner radius of the accretion disk is equal or larger than the corotation radius (i.e. when $\rm R_t \geq R_{co}$). In such a situation the spin-down efficiency of the magnetospheric ejections is maximized since they can be directly powered and accelerated by the stellar rotation, and it becomes more important than the torque exerted by the stellar wind. 

On the other hand, the centrifugal barrier produced by the stellar rotation makes accretion more difficult since the inner edge of the disk tends to be spun-up by the stellar rotation. Typically, in this regime accretion occurs in cycles \citep{Romanova05, Lii14, Romanova18}, which determine a strong variability (no accretion to accretion bursts) and occur on relatively short timescales (a few stellar periods). It is important to notice that this extreme variability could already occur when the disk is truncated slightly inside the corotation radius.\ This is due to the fact that as the disk tends to rotate at a sub-Keplerian rate in the truncation region, it can already feel the stellar rotational barrier when $\rm R_t \lesssim R_{co}$ \citep{Zanni2013}. As far as we know, this kind of variability has never been observed in CTTS. It must be pointed out that the axisymmetric MHD simulations commonly used to investigate the propeller regime could strongly amplify this effect. As a matter of fact, AA Tau, which we recall being the only star in Table \ref{obs} to be efficiently spun-down when applying our $\rm Q_{acc} = 1\%$ model, should be truncated very close to the corotation radius, but this star has never shown the variability usually found in propeller models. Both conditions are theoretically (the mass carried by stellar wind i.e. $\rm Q_{acc}=10\%$) and observationally (the strong variability introduced by the propeller regime) problematic.

\subsection{Early-PMS rotation rate}

Since the results of the models are compared to the observations, the cluster's age is a key parameter. Unfortunately, the age of PMS clusters is poorly constrained yet \citep{Bell13} this factor limits the strength of the results presented here. The uncertainties in the rotation period measurement induced by, for example, synchronized binaries and multiple spots at the stellar surface \citep{Bouvier97,Moraux13} are in principle already included in the error bars given by the rejection method used in this study \citep[see][]{Gallet13,GB15}. Moreover, the age estimations come from different observations, methods, and techniques that thus provide an inhomogeneous temporal sample. This highlights the need for a self-consistent analysis of clusters properties.

Besides this age estimation limitation, there are several observational biases in terms of rotation period measurement. These measurements are usually realized by extracting quasi-periodic modulations in the photometric observation of a given star that is induced by the presence of surface stellar spots. Hence, the nature of the technique itself leads to observations that are more sensitive to magnetically active fast rotating stars, for which several complete rotational periods can be monitored during the observation time.

We can also mention that the metallicity could have a strong effect on the rotational evolution of low-mass stars, even during their early-PMS phase evolution. Indeed, changing the initial metal content of a star can mimic the effect of changing its initial mass on the evolution of the stellar internal structure. A decrease of metallicity induces a reduction of the global opacity of the star that allows for an easier redistribution of the energy inside of the convective envelope and thus leads to an increase of the stellar surface effective temperature \citep[see][]{Amard19}.

\subsection{{Observational evidences of magnetopsheric outflows}}

{Outflows from the innermost parts of the star-disk interaction region play a fundamental role for the stellar spin evolution in our framework. They have been typically associated with the blue-shifted components either in the emission of forbidden lines, or in the absorption of permitted ones.} 

{For example the P-Cygni profile of the He I $\lambda$10830 line has proven to be a sensible diagnostic for both infall and outflow \citep{Dupree2005, Edwards06}. In particular \cite{Edwards06} propose that broad and deep blue-shifted absorption features could be associated with a stellar wind in systems seen at low inclinations, as in the TW Hya case \citep{Dupree14}. Radiative transfer calculations have confirmed this hypothesis but, at least in the specific case of AS 353 A, a massive ($\rm \dot{M}_{wind} > 10^{-9} \, M_\odot yr^{-1}$) and relatively cold ($8000$ K) wind is required to reproduce the observations \citep{Kurosawa11}. These estimates raise questions about the nature of these putative stellar winds, as previously mentioned in Section \ref{interregime}.}     

{On the other hand, \cite{Edwards06} associate narrow blue-shifted absorption features of the He I $\lambda$10830 line with outflows emerging from the inner disk of systems seen at high inclinations. In the case of AA Tau, a prototypical CTTS seen at high inclination, \cite{Bouvier03} observed a correlation between the variations of the radial velocity of the blue-shifted (outflow) and red-shifted (inflow) absorption components in the H$\alpha$ line profile. The authors interpret this correlation to be due to a periodic inflation and reconnection of the stellar magnetic field. This interpretation is qualitatively consistent with the behaviour of magnetospheric ejections, but the variability that the numerical models of \cite{Zanni2013} require to efficiently slow down the stellar rotation is much more extreme than the one displayed, for example, by AA Tau. In general, young stellar objects are known to be characterized by accretion and ejection outbursts \cite[see e.g.][]{Ansdell16} but, as far as we know, this burst-like behaviour has never been observed on the relatively short timescales (a few stellar periods) that seem to characterize the simulations of the propeller regime.} 

{In any case, it is not clear whether it is possible or not to use the blueshifted absorption components to differentiate between the possible inner disk outflow scenarios, that is, MEs, an X-wind, an ReX-wind, or the inner part of a more extended disk-wind.
For example the radiative transfer calculations of \cite{Kurosawa11} and \cite{Kurosawa12} that are based on a semi-analytic toy model of a more extended disk-wind and a numerical MHD model of a conical wind solution (that we think to be closely related to the MEs), respectively, provide qualitatively the same line profiles. More detailed radiative transfer calculations of different outflow classes are required to attempt a more quantitative comparison with observations and constrain the properties of these winds and their possible influence on the stellar spin evolution.}

\section{Summary and conclusions}
\label{conc}

Different measurements of the rotational period distribution of young star-forming regions lead to the conclusion that the surface rotation rate of stars remains approximately constant during their early-PMS phase. {This stage seems to be related to the epoch during which the forming stars are still surrounded by, and interact with, an accretion disk.} These observations hence have motivated angular momentum {evolution models} to often simply consider a constant surface angular velocity during the first few Myr of the stellar life so as to mimic this observed feature \citep{Gallet13,GB15,Amard16}.

{To improve this simplified vision of a constant surface rotation rate, we decided to include an actual star-disk interaction model in our angular momentum evolution calculations so as to investigate what properties of CTTS are required to fulfill the observations constraints.}
In this study, we directly compared the angular velocity evolution {that results from} {our} star-disk interaction {model} to {rotation period} observations
{of solar-type stars}. 

We pointed out that {a kG} dipolar magnetic field component is {typically} required during the entire disk lifetime so as to extract enough angular momentum from the stellar surface to compensate {for} the acceleration of the stars due to their contraction {and accretion}. 
{While such strong dipolar magnetic field intensity is sometimes detected, it is not ubiquitous. Indeed, at the very beginning of their evolution, young and fully convective CTTS (e.g. AA Tau and BP Tau) {are sometimes observed} to display strong dipolar magnetic field components between 1 kG and 2 kG. This dipolar magnetic field intensity is then ideal for rotational regulation through star-disk magnetic interaction processes.} 

{Besides, we find that, to have an efficient spin-down, the interaction regimes are often rather extreme. Our models frequently require either of the following: firstly, strong stellar winds, with a mass loss rate around 10\% of the accretion rate, that {seem hard to produce }due to general energetic considerations; or secondly, being in a propeller regime ($\rm R_t > R_{co}$) that maximizes the spin-down efficiency of magnetospheric ejections, but at least according to axisymmetric numerical models, triggers an extreme accretion variability that is generally not observed.}

The results of this work should, however, be considered as preliminary and a more physical model has yet to be developed. More specifically, we should investigate star-disk interaction models in which the impact of a non-axisymmetric and multipolar magnetic field is taken into account, or include other effects that could influence the spin evolution, such as the accretion and ejection bursts associated with FU Ori events that are neglected in this work. 

\begin{acknowledgements}
{We thank the anonymous referee for the constructive comments about our work that increased the quality of the paper.} F.G acknowledges financial support from the CNES fellowship. This study was supported by the grant ANR 2011 Blanc SIMI5-6 020 01 `Toupies: Towards understanding the spin evolution of stars' (\url{http://ipag.osug.fr/Anr_Toupies/}). We acknowledge financial support from CNRS-INSU's Programme National de Physique Stellaire. This project has received funding from the European Research Council (ERC) under the European Union's Horizon 2020 research and innovation program (grant agreement No 742095; {\it SPIDI}: Star-Planets-Inner Disk-Interactions). F.G acknowledges financial support from the CNES fellowship. {LA gratefully acknowledges financial support from the ERC (grant 682393, {\it AWESoMe Stars}).} We thank C. Johnstone for enlightening discussions on CTTS magnetic field evolution.
\end{acknowledgements}

\bibliographystyle{aa}
\bibliography{Bib}

\end{document}